\documentstyle[aps,multicol,epsfig]{revtex}
\begin{document}
\draft
\title{Scaling Behavior of Anomalous Hall Effect and Longitudinal
Nonlinear Response in High-$T_{C}$  Superconductors
\thanks{This work is supported by the Ministry of Science
and Technology of China (NKBRSF-G 19990640) and the Chinese
NSF.}}
\author{H.Y.Xu, Z.Qi, W.Wang, D.L.Yin$^\ast$, F.R.Wang, and C.Y.Li}
\address{Department of Physics, Peking University, Beijing 100871, China}
\maketitle

\begin{abstract}
Based on existing theoretical model and by considering our
longitudinal nonlinear response function, we derive a nonlinear
equation in which the mixed state Hall resistivity can be
expressed as an analytical function of magnetic field,
temperature and applied current. This equation enables one to
compare quantitatively the experimental data with theoretical
model. We also find some new scaling relations of the temperature
and field dependency of Hall resistivity $\rho_{x\!y}$. The
comparison between our theoretical curves and experimental data
shows a fair agreement.

\vspace{0.5cm}
\end{abstract}


\widetext

\begin{multicols}{2}

{\it Introduction}.--The anomalous Hall effect (AHE) remains one
of most intriguing phenomena in the area of high-temperature
superconductors (HTS). A sign reversal of the Hall resistivity
$\rho_{x\!y}$ has been observed in the most of HTS just below the
transition temperature $T_{C}$ \cite{R1} although the conventional
theories of flux motion proposed by Bardeen and Stephen \cite{R2} as
well as Nozi\`{e}res and Vinen \cite{R3} predict that the Hall effect
stems from quasi-normal core and hence has the same sign as in the
normal state. Several attempts at a theoretical understanding of
this surprising sign reversal have been undertaken \cite{R4}. Taking
into account the backflow current due to pinning and thermal
fluctuation effect, Wang, Dong and Ting (WDT) \cite{R5} developed a
unified theory for the anomalous sign reversal and the observed
puzzling scaling relation between the Hall and longitudinal
resistivities $\rho_{x\!y}\propto\rho_{x\!x}^\beta$ \cite{R6}. They
obtained a relation:
\begin{equation}
\rho_{x\!y}=\frac{\beta_{0}\rho_{x\!x}^{2}}{\Phi_{0}B}\{\eta(1-\overline{\gamma})-2\overline{\gamma}\Gamma(\upsilon_{L})\}
\end{equation}
      where $\beta_{0}=\mu_{m}H_{C\!2}$ with $\mu_{m}=\tau e/m$ the mobility of the charge carrier and $H_{C\!2}=\Phi_{0}/2\pi\xi^{2}$ being the usual upper critical
field with $\xi$ the superconducting coherence length, and $\eta$
is the viscous cofficient.
$\overline{\gamma}=\gamma(1-\overline{H}/H_{C\!2})$ with
$\overline{H}$ the average magnetic field over core and $\gamma$
the parameter describing contact force on the surface of core. $B$
is the magnetic field. $\Gamma(\upsilon_{L})$ is a positive scale
function between the time average pining force $\langle {\bf
F}_{p}\rangle_{t}$ and flux motion velocity
$\upsilon_{L}\equiv\langle\upsilon_{\Phi(i)}\rangle_{t}$ defined
as:
\begin{equation}
\langle{\mathbf F}_{p}\rangle_{t}=-\Gamma(\upsilon_{L}){\bf v}_{L}
\end{equation}
However, as mentioned by WDT, it is difficult to find analytical
expression for $\Gamma(\upsilon_{L})$ theoretically though two
approximate methods were suggested.

Another school of thought considers the intrinsic vortex
properties. Vinokur, Geshkenbein, Feigel'man and Blatter (VGFB)
\cite{R7} explain the observed scaling behavior
$\rho_{x\!y}\propto\rho_{x\!x}^{2}$ assuming that the Hall
conductivity $\sigma_{x\!y}$ does not depend on disorder. A
microscopic analysis of intrinsic dynamical single vortex
properties provides an interpretation of the observed sign change
in the Hall effect in the superconductors with mean free path
$\ell$ of the order of coherence length $\xi$ in terms of broken
particle-hole symmetry, which is related to detail of the
microscopic mechanism of superconductivity.

For testing the role of pinning and intrinsic vortex properties on
the Hall effect, experiments to measure the $\rho_{x\!y}$ or
$\sigma_{x\!y}$ before and after heavy-ion irradiation were
carried on since heavy-ion track are very effective pinning
centers \cite{R8}. Another way to vary the effect of pinning is
measuring $\rho_{x\!y}$ at different current density $J$ \cite{R9}. In
spite of much recent efforts, qualitative comparison between
theories and experimental data remains controversial and
consensus on the origin of AHE is still lacking.

In present work, we try to make a rather quantitative comparison
between experimental data and theoretical models. By using a
recently found nonlinear material equation \cite{R10,R11}, we derive a
theoretical expression for $\rho_{x\!y}$ which is a function of
temperature, applied magnetic field and applied current. We also
find some new scaling relations of the temperature and field
dependency of Hall resistivity $\rho_{x\!y}$. Comparison between
the  scaled experimental data and our theoretical curves for
$\rho_{x\!y}$ showed fair agreement.

{\it Hall resistivity equation}.--In the mixed state of type-II
superconductors, flux enters the superconductors in form of
quantized flux lines or vortices according to the theory of
Abrikosov. In the case of nonideal type-II superconductors the
flux lines are pinned by imperfections, small current flow usually
results in vanishing resistance. However, with a sufficiently
large driving force produced by a significant transport current,
flux lines can be depinned. Their motion is subject to a viscous
drag, giving rise to dissipation \cite{R2}. Therefore, for the steady
state of flux motion in nonideal type-II superconductors, the mean
current transport density $J$ can be phenomenologically expressed
as a sum
\begin{equation}
J=J_p+J_f
\end{equation}
with $J_f\equiv E(J)/\rho_f$ the component due to the moving
vortices of uniform density and $J_p$ denotes contribution from
the pinned vortices with nonuniform distribution. Therefore, the
pinning force $\langle{\mathbf F}_p\rangle_{t}$ can be expressed
as
\begin{equation}
\langle{\mathbf F}_p\rangle_{t}=-{\mathbf J}_{p}\times {\mathbf B}
\end{equation}
and we have
\begin{equation}
\Gamma(\upsilon_L)=\frac{(J-J_{f})B^{2}}{E}
\end{equation}
In writing this, we have used Maxwell equation ${\mathbf E=B\times
v}_{L}$ with ${\mathbf v}_L$ flux motion velocity which induces an
electric field ${\mathbf E}$. Substituting the right hand side of
Eq.(5) into the term $\Gamma(\upsilon_L)$ in Eq.(1) and following
the estimation in Ref.[2] as $\eta=B^{2}/\rho_{f}$ and the
resistivity due to flux flow
$\rho_{f}=E/J_{f}=\rho_{n}B/B_{C\!2}$, we get
\begin{equation}
\rho_{x\!y}=\frac{\beta_{0}
B^{2}\rho_n}{\Phi_{0}B_{C\!2}}\{(1+\overline{\gamma})\frac{\rho_{x\!x}^{2}}{\rho_{f}^2}
-2\overline{\gamma}\frac{\rho_{x\!x}}{\rho_f}\}
\end{equation}
where $\rho_n$ and $B_{C\!2}$ are the resistivity of normal state
and the upper critical magnetic field, respectively.

The form of Eq.(6) implies that the behavior of $\rho_{x\!y}$
strongly depends on the longitudinal $\rho_{x\!x}$, i.e., the
longitudinal nonlinear response of HTS, which will lead to
nonlinear Hall resistivity and even sign reversal.
P.J.M.W\"{o}ltgens {\it et al.} \cite{R12} measured the nonlinear Hall
resistivity in $YB_{2}Cu_{3}O_{7-\delta}$ film near the vortex-glass
transition. Their results firmly indicated that the Hall
resistivity has the similar scaling relation as for longitudinal
resistivity which has been discussed extensively by M.P.A.Fish
{\it et al.} \cite{R13}. Therefore the main problem for the Hall
resistivity equation is to find out a general analytical
expression of the nonlinear longitudinal $\rho_{x\!x}$ usually
formulated in the form
\begin{equation}
\rho_{x\!x}=E(J)/J=\rho_{f}e^{-U(T,B,J)/kT}
\end{equation}
in which different types of the current dependency $U(J)$ have
been suggested to approximate the real barrier, for instance, the
Anderson-Kim model \cite{R14} with $U(J)=U_{C}(1-J/J_{C\!0})$, the
logarithmic barrier $U(J)=U_{C}\ln(J_{C\!0}/J)$ \cite{R15} and the
inverse power-law with $U(J)=U_{C}[(J_{C\!0}/J)^{\mu}-1]$ \cite{R16}.

We find, if one makes a common modification to the different
model barriers $U(J)$ as
\begin{equation}
U(J)\longrightarrow U(J_{p}\equiv J-E/\rho_{f})
\end{equation}
the corresponding modified material equation
\begin{equation}
E(J)=J\rho_{f}e^{-U(J_{p})/kT}
\end{equation}
leads to a common form of longitudinal resistivity [10,11]

\begin{equation}
\rho_{x\!x}=\rho_{f}\exp[{-a\frac{U_C}{kT}\cdot J^p\cdot
(\frac{1}{J}+\frac{\rho_{x\!x}/\rho_{f}}{J_{C\!0}}-\frac{1}{J_{C\!0}})^p
}]
\end{equation}
where $U_{C}$ and $J_{C\!0}$ are only dependent on the temperature
and magnetic field. Following the discussion in the review article
of Cohen {\it et al.} \cite{R15}, we can assume that the temperature and
field are separated variables in the $U_C$ and $J_{C\!0}$ and
write
\begin{equation}
U_C\propto
B^{m}(1-\frac{B}{B_{C\!2}})^{\delta}(1-\frac{T}{T^\ast})^\alpha
\end{equation}
\begin{equation}
J_{C\!0}\propto (1-\frac{T}{T^\ast})^\beta
\end{equation}
with $T^\ast$ the irreversibility temperature and $m$, $\delta$,
$\alpha$ and $\beta$ are exponents. Combining Eq.(6) with
Eq.(10-12), we finally get the analytical expression for Hall
resistivity
\end{multicols}
\linethickness{0.7pt}
\begin{picture}(4,5)
\put(-5,4){\line(1,0){90}}
\put(85,3.9){\line(0,1){2}}
\end{picture}
\begin{eqnarray}
\nonumber\rho_{x\!y}=\frac{\beta_{0}ATB^{2}}{\Phi_{0}B_{C\!2}}\{(1+\overline{\gamma})\exp[{-2a\frac{B^{m}(1-\frac{B}{B_{c\!2}})^{\delta}(1-\frac{T}{T^\ast})^\alpha}{kT}J^{p}(\frac{1}{J}+\frac{\rho_{x\!x}/\rho_{f}}{b(1-\frac{T}{T^\ast})^\beta}-\frac{1}{b(1-\frac{T}{T^\ast})^\beta})^{p}}]\\
-2\overline{\gamma}\exp[-a\frac{B^{m}(1-\frac{B}{B_{C\!2}})^{\delta}(1-\frac{T}{T^\ast})^\alpha}{kT}J^{p}(\frac{1}{J}+\frac{\rho_{x\!x}/\rho_{f}}{b(1-\frac{T}{T^\ast})^\beta}-\frac{1}{b(1-\frac{T}{T^\ast})^\beta})^{p}]
\}
\end{eqnarray}
\begin{picture}(4,5)
\put(90,0){\line(1,0){90}}
\put(90,0.2){\line(0,-1){2}}
\end{picture}

\begin{multicols}{2}
where we have used the approximate resistivity relation
$\rho_{n}=AT$ for HTS with $A$ as a constant. In Fig.1, we plot a
series of $\rho_{x\!y}$ vs. $B$ curves as the interesting
numerical solutions of Eq.(13) with the values of exponents of
$m$, $\delta$, $\alpha$, $\beta$ similar to that of Ref.\cite{R17} and
different values of $\gamma$ which describes contact force on the
surface of core. Negative Hall resistivity appears between two
characteristic fields, $B_r$ (where sign reversal occurs) and
$B_0$ (where $\rho_{x\!y}$ disappears), or between two similar
characteristic temperature $T_r$ and $T_0$ in $\rho_{x\!y}\sim T$.
It is also clearly seen that the magnitude of negative Hall
resistivity decreases as the value of $\gamma$ decreasing and
vanishes at a definite value. This result coincides with the
conclusion of WDT \cite{R5}. Moreover, it is worthy to note that, some
times the numerical solution of $\rho_{x\!y}\sim B$ (see Fig.1)
manifests two sign reversals of Hall resistivity $\rho_{x\!y}$,
i.e., the Hall resistivity changes sign from negative to positive
at a lower magnetic field besides the first change from positive
to negative at higher field. This behavior is very similar to that
experimentally observed by J.Schoenes et al. \cite{R18}.

{\it Scaling behavior of AHE}.--Besides the well known scaling
relations between Hall and longitudinal resistivities
$\rho_{x\!y}\propto (\rho_{x\!x})^{\beta}$, we find two further
striking scaling relations of AHE as
\begin{equation}
\frac{\rho_{x\!y}}{\rho_{m}} \approx f
(\frac{B-B_{0}}{B_{r}-B_{0}})
\end{equation}
\begin{equation} \frac{\rho_{x\!y}}{\rho_{m}}\approx
f'(\frac{T-T_0}{T_{r}-T_{0}})
\end{equation}
where $\rho_{m}$ is the maximal negative Hall resistivity, $f$
and $f'$ are scaling functions which in general can be dependent
on the specific samples measured in the experiments. However, to
our astonishment, we find that the scaled experimental data
obtained from different laboratories (Ref.\cite{R9}, Ref.\cite{R19} and
Ref.\cite{R20}) also fall onto a single scaling curve (see Fig.2). With
the similar scaling Eq.(15), the negative Hall resistivity data
from S.J.Hagen {\it et al.} \cite{R21} and Y.Matsuda {\it et al.} \cite{R22},
measured at many different magnetic fields can also collapse onto
a single universal functional dependence on the scaled
temperature as shown in Fig.3. For making a comparison between
the experimental data and theoretical predication, it is
convenient to rescale the curves obtained from theoretical
expression of $\rho_{x\!y}$ according to Eq.(14) and Eq.(15).
Fig.2 and Fig.3 show comparison between the scaled experimental
data and the scaled theoretical curves with the values of
exponents of $m=2.0$, $\delta=4.2$, $\alpha=1.8$, $\beta=1.8$,
$p=1$ as used in Ref.\cite{R17}. We see fair agreements for both the
scaled field and temperature dependence of $\rho_{x\!y}/\rho_{m}$.

{\it Discussion}.--Though the comparisons of our Hall resistivity
equation (see Eq.(13)) with the pertinent experimental results
shown in Fig.2 and Fig.3 convincingly indicate the pinning
dependence of $\rho_{x\!y}$, it by no means rules out the
possible role of particle-hole asymmetry in AHE \cite{R4}. Actually,
Eq.(13) is not incompatible with a possible broken particle-hole
symmetry in the electronic band structure of superconductor.
However, we believe the rather quantitative fit of experimental
data of different origins with our Hall resistivity equation in
the scaling form Eq.(14)(15) may imply that pinning is the key
factor in AHE of high-$T_{C}$ cuprates. It would be very
interesting to see whether the anomalous Hall effect is still
observed in the recently discovered high-$T_{C}$ ($\sim 40K$)
$MgB_{2}$ \cite{R23} which in contrast to the high-$T_{C}$ cuprates
manifests typical electron-phonon superconductivity. It is also
useful to check the $\rho_{x\!y}$ scaling behavior in this
conventional s-p band metallic compound and compare it with those
for the high-$T_{C}$ cuprates as strongly correlated electronic
systems.

{\it Summary}.--Based on the WDT's model and combining our
nonlinear response equation, we derived a theoretical equation
for mixed state Hall resistivity in type-II superconductors. This
equation predicts the sign reversal of Hall resistivity and has a
quantitative agreement with experimental data.

{\it Note added in proof} : After preparation of this paper we
became aware of the interesting work of R.Jin {\it et al.} on the
Hall effect in $MgB_2$ film \cite{R24}. The experimentally observed
$R_{H}\sim B$ and $R_{H}\sim T$ data clearly indicated that there
are sign reversals of Hall resistivity in $MgB_2$ very similar to
our Eq.(14) and Eq.(15) and the behaviors shown in Fig.2 and
Fig.3.

\begin{figure}[F1]
\epsfxsize= .83\hsize  \vskip 1.0\baselineskip \centerline{
\epsffile{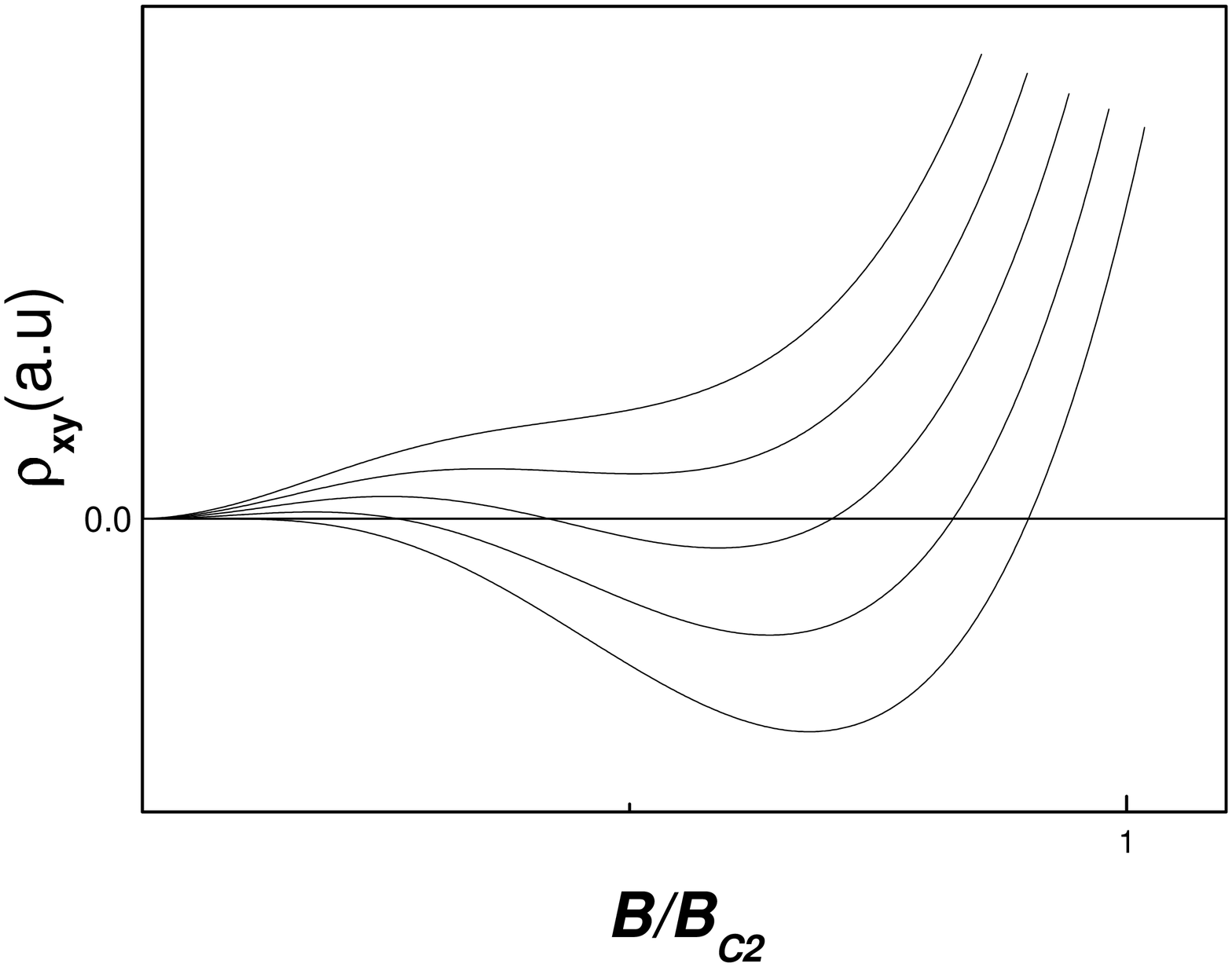}}
 \vspace{.1cm}
\caption{ The theoretical
curves of $\rho_{xy}$ as a function of reduced field $B/B_{C2}$
for several value of the parameter $\gamma$ ($0<\gamma<1$). The
values of $\gamma$ from bottom to top are 0.8, 0.7, 0.6, 0.5, and
0.4.} 
\end{figure}
 \vspace{-.8cm}

 \begin{figure}[F2]
\epsfxsize= .83\hsize  \vskip 1.0\baselineskip \centerline{
\epsffile{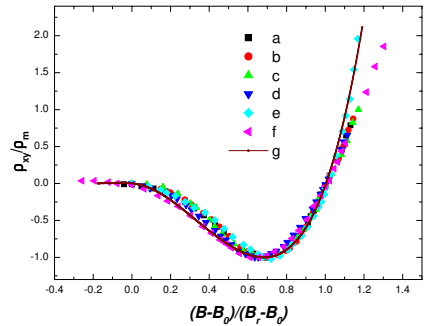}}
 \vspace{.1cm}
\caption{ Empirical scaling
functions for $\rho_{x\!y}$ vs. scaled magnetic field for various
values of temperature. The experimental data (a)-(d) are from
A.W.Smith {\it et al.} (Ref.[9] Fig.1) and (e), (f) are from J.Luo
{\it et al.} (Ref.[19] Fig.1) and M.R.Cimberle {\it et al.}
(Ref.[20]), respectively. The solid curve (g) is theoretical
result obtained from Eq.(13). The comparison between the
theoretical curve and experimental data shows a fair agreement.}
\end{figure}
 \vspace{-.8cm}

 \begin{figure}[F3]
\epsfxsize= .83\hsize  \vskip 1.0\baselineskip \centerline{
\epsffile{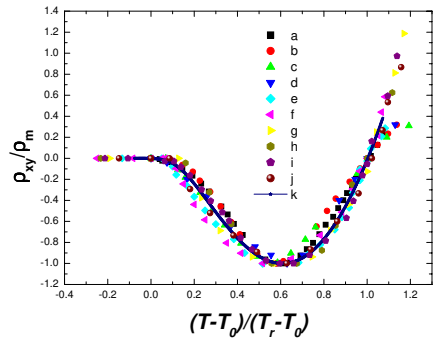}}
 \vspace{.1cm}
\caption{Empirical scaling
functions for $\rho_{x\!y}$ vs scaled temperature for various
values of magnetic fields. The experimental data (a)-(d) are from
S.J.Hagen {\it et al.} (Ref.[21] Fig.1) and (e)-(j) are from
Y.Matsuda {\it et al.} (Ref.[22] Fig.2 and Fig.3). The solid curve
(k) is theoretical result obtained from Eq.(13). The comparison
between the theoretical curve and experimental data shows a fair
agreement.}
 \end{figure}

\end{multicols}

\end{document}